\documentclass[12pt]{article}

\catcode`\@=11
\@addtoreset{equation}{section}

\global\arraycolsep=2pt
\usepackage{amsbsy,amssymb,latexsym,amsfonts,amsmath}
\usepackage{graphicx,color}

\allowdisplaybreaks

\begin{document}

\begin{flushright}
\parbox{4.2cm}
{RUP-21-4}
\end{flushright}

\vspace*{0.7cm}

\begin{center}
{ \Large  On an alternative quantization of R-NS strings}
\vspace*{1.5cm}\\
{Tsubasa Yuki and Yu Nakayama}
\end{center}
\vspace*{1.0cm}
\begin{center}

Department of Physics, Rikkyo University, Toshima, Tokyo 171-8501, Japan

\vspace{3.8cm}
\end{center}

\begin{abstract}
We investigate an alternative quantization of R-NS string theory. In the alternative quantization, we define the distinct vacuum for the left-moving mode and the right-moving mode by exchanging the role of creation operators and annihilation operators in the left-moving sector. The resulting string theory has only a finite number of propagating degrees of freedom. We show that an appropriate choice of the GSO projection makes the theory tachyon free. The spectrum coincides with the massless sector of type IIA or type IIB superstring theory without any massive excitations. 
\end{abstract}

\thispagestyle{empty} 

\setcounter{page}{0}

\newpage

\section{Introduction}

The birth of the string theory was between the late 1960s and early 70s \cite{Nambu}. At the time, they came up with the idea that some properties of hadrons can be explained from a picture of strings, and string theory may be able to describe the strong interaction among hadrons. They soon realized that there are certain problems in the quantization of strings in four-dimensions. Furthermore the advent of QCD that describes the strong interaction as a gauge theory made string theory almost forgotten.

Around 1974, however, Yoneya \cite{Yoneya:1973ca}  as well as Scherk and Schwarz  \cite{Scherk:1974ca} proposed that we should regard massless spin-two excitations in string theory as a graviton rather than hadrons. This change of view has made the string theory a candidate for a consistent quantum theory of gravity without ultraviolet divergence. Still, the bosonic string theory has problems such as the existence of tachyon.

Then superstring theory that circumvents these problems was discovered.  In 1971, Ramond introduced the stringy version of the Dirac equation \cite{Ramond:1971gb}. Later, Neveu and Schwarz studied a quantization of the same theory with a different boundary condition \cite{Neveu:1971rx}. They are now known as R-NS formulation of the superstring theory. In 1972, Schwarz discovered that the critical dimension of the R-NS string theory is ten dimensions \cite{Schwarz:1974ix}, and in 1977, Gliozzi, Scherk, and Olive introduced the so-called GSO projection \cite{Gliozzi:1976qd} on the spectrum of the R-NS string theory. This projection removes the tachyon from the physical spectrum and results in the target-space supersymmetry. 

Recently, there have been some attempts to quantize the string theory in an alternative way. In conventional quantization of the (bosonic) string theory, the world-sheet vacuum is symmetric under the exchange of the left-moving mode and the right-moving mode. In \cite{Lee:2017utr}, the alternative quantization by reversing the creation and annihilation of the left-moving mode was proposed.\footnote{There are a couple of earlier works on the asymmetric vacuum choice in the tensionless string theories \cite{Gamboa:1989px}\cite{Gamboa:1989zc}\cite{Casali:2016atr}. As we will see, the resulting string theory is more or less equivalent to the chiral string theory proposed in \cite{Siegel:2015axg}\cite{Huang:2016bdd}\cite{Leite:2016fno} (see also \cite{Jusinskas:2019dpc}). The alternative choice of the vacuum is called twisted vacuum and the resulting theory is called twisted strings in \cite{Casali:2016atr}\cite{Casali:2017mss}.} 
The resulting theory has a very peculiar feature that only a finite number of degrees of freedom propagate in the space-time. The aim of this paper\footnote{It is based on the first author's master thesis originally written in Japanese. It is  edited and translated by the second author. Compared with the original thesis, the BRST quantization of the bosonic string theory with the alternative vacuum choice and a couple of appendices are omitted while some footnotes are added.} is to investigate the alternative quantization in the R-NS string theory. 

The organization of the paper is as follows. In section 2, we review the alternative quantization of bosonic string theory. In section 3, we investigate the alternative quantization of the R-NS string theory with and without GSO projection. Section 4 is devoted to discussions.

\section{Alternative quantization of bosonic string theory}
We first consider a bosonic string theory propagating in the $D$-dimnsional Minkowski space-time equipped with the Cartesian coordinate $X^\mu$. We use the Lorentz invariant inner-product $A\cdot B = \eta_{\mu\nu}A^\mu B^\nu$ with $\eta_{\mu\nu} = \text{diag}(-1,1, \cdots, 1)$. The worldsheet of the string  is parametrized by $\sigma^0 = \tau$ and $\sigma^1 = \sigma$. We focus on the closed string theory with a periodic boundary condition $\sigma \sim \sigma + 2\pi$.

In the conformal gauge, the worldsheet closed string mode $X^\mu$ is described by the Polyakov action
\begin{align}
S = \frac{1}{4\pi \alpha'} \int d^2\sigma \left( \partial_\tau X \cdot \partial_\tau X  - \partial_{\sigma} X \cdot \partial_\sigma X \right) \ 
\end{align}
 with the equation of motion
\begin{align}
(-\partial_\tau^2 + \partial_\sigma^2) X^\mu = 0 \ .
\end{align} 
The general solution of the equation of motion is given by a sum of a left-moving mode $X^\mu_L$ and a right-moving mode $X^\mu_R$:
\begin{align}
X^\mu(\tau,\sigma) = X_R^\mu (\tau- \sigma) + X_L^\mu (\tau + \sigma) \ . 
\end{align}

With the closed string boundary condition
\begin{align}
X^\mu (\tau,\sigma) = X^\mu (\tau, \sigma + 2\pi)
\end{align}
the mode expansion of $X^\mu$ is explicitly given by
\begin{align}
X_R^\mu &= \frac{1}{2} x_R^\mu + \alpha' \frac{p^\mu_R}{2} (\tau-\sigma) + i \sqrt{\frac{\alpha'}{2}} \sum_{n\neq 0} \frac{1}{n} \alpha^\mu_n e^{-2in (\tau-\sigma)} \cr
X_L^\mu &= \frac{1}{2} x_L^\mu  + \alpha' \frac{p^\mu_L}{2} (\tau+\sigma) + i \sqrt{\frac{\alpha'}{2}} \sum_{n  \neq 0} \frac{1}{n} \bar{\alpha}^\mu_n e^{-2in (\tau+\sigma)} \ .
\end{align}
In the conventional quantization of closed string theory, we assume $x_R^\mu = x_L^\mu \equiv x^\mu $ and $p_R^\mu = p_L^\mu \equiv p^\mu$. 

In the alternative quantization proposed in \cite{Casali:2016atr}\cite{Lee:2017utr}, we are going to choose an asymmetric vacuum for the left-moving mode and right-moving mode, so we have to treat the left-right zero modes independently. Accordingly we  postulate the commutation relation 
\begin{align}
[x_R^\mu, p_R^\nu] = i \eta^{\mu\nu} \ , \ \ [x_L^\mu, p_L^\nu] = i \eta^{\mu\nu} \ . 
\end{align}
Otherwise, they satisfy the same commutation relation as in the conventional quantization of the closed string theory:
\begin{align}
[\alpha_m^\mu, \alpha_n^\nu] = [\bar{\alpha}^\mu_m, \bar{\alpha}^\nu_n] = m \delta_{m+n, 0} \eta^{\mu\nu} \ . 
\end{align}

Following \cite{Lee:2017utr}, we now define the alternative vacuum for the left-moving mode and the right-moving mode as
\begin{align}
\alpha_n^\mu | 0, k \rangle &= 0 \ , \ \  \overline{\langle 0, k|} \bar{\alpha}^\mu_n = 0 \cr
p_R^\mu |0, k \rangle & = k^\mu |0, k \rangle  \ , \ \  \overline{\langle 0, k|} p_L^\mu =  \overline{\langle 0, k|} k^\mu \  
\end{align}
for $n>0$.\footnote{The condition $k_L^\mu  = k_R^\mu$ must be regarded as a constraint on the states. This makes the situation more complicated if we compactify the target space as discussed in \cite{Lee:2017crr}\cite{Casali:2017mss}.} The state with a bar emphasizes that we use the alternative vacuum choice for the left-moving mode.

In order to analyze the closed string spectrum, we introduce the level operator. The level operator for the right-moving mode is same as the one in the conventional quantization of the bosonic string theory:
\begin{align}
N = \sum_{n=1}^\infty \alpha_{-n} \cdot \alpha_n \ .
\end{align}
For the left-moving mode, we, however, define the level operator as
\begin{align}
\bar{N} = - \sum_{n=1}^\infty \bar{\alpha}_n \cdot \bar{\alpha}_{-n} 
\end{align}
so that it becomes semi-positive definite on excited states of our alternative vacuum. 

With this definition of the level operators, zero modes of the Virasoro generator are expressed as\footnote{We did not include the normal ordering constant here.}
\begin{align}
L_0 = \frac{1}{2}\alpha_0^2 + N \ , \ \ \bar{L}_0 = \frac{1}{2}\alpha_0^2 - \bar{N} \ , 
\end{align}
where $\alpha_0^\mu = \sqrt{\frac{\alpha'}{2}} p_R^\mu \ , \bar{\alpha}_0^\mu = \sqrt{\frac{\alpha'}{2}} p_L^\mu$.
 The definition of the other modes of the Virasoro generators is 
\begin{align}
L_n = \frac{1}{2} \sum_m  \alpha_{n-m} \cdot \alpha_m  \ , \ \ \bar{L}_n = \frac{1}{2} \sum_m  \bar{\alpha}_{n-m} \cdot \bar{\alpha}_m   \ . 
\end{align}

In the classical Polyakov formulation of the bosonic string theory, we need to impose the constraint on the energy-momentum tensor from the variation of the worldsheet metric:
\begin{align}
T_{\alpha\beta} = 0 \ . 
\end{align}
In the (old) covariant formulation of the alternative quantization of the bosonic string theory, we interpret it as 
\begin{align}
L_m | \text{phys} \rangle = 0 \ , \ \  \langle \text{{phys}}| \bar{L}_m = 0 
\end{align}
for $m>0$,
and
\begin{align}
(L_0 - a) | \text{phys} \rangle = 0 \ , \ \  \langle \text{{phys}}| (\bar{L}_0 - \bar{a}) \ ,
\end{align}
where $a$ and $\bar{a}$ are normal ordering constants to be determined.
 The zero mode conditions are equivalent to the mass-shell condition
\begin{align}
M^2 = \frac{4}{\alpha'} ( N -a)  = \frac{4}{\alpha'} (-\bar{N} - \bar{a}) \label{os}
\end{align}
and the level-matching condition
\begin{align}
 N + \bar{N} = a - \bar{a} \ . \label{lm}
\end{align}
It is crucial to observe the sign difference in front of $\bar{N}$  in \eqref{os} and \eqref{lm} compared with the conventional quantization of the bosonic string theory.

Following \cite{Lee:2017utr}, we assume that the level one states including graviton will be in the physical states and massless. This determines the normal ordering constant as $a=1$ and $\bar{a}= -1$. It also fixes the space-time dimension $D=26$. Once $a$ and $\bar{a}$ are determined, the mass-shell condition and the level-matching condition become
\begin{align}
M^2 &= \frac{4}{\alpha'} (N-1) = \frac{4}{\alpha'} ( - \bar{N} +1) \cr
N + \bar{N} & = 2 \ . 
\end{align}

In the conventional quantization of the closed string theory, the level-matching condition is $N = \bar{N}$ and the infinite excited modes are in the physical states, but in the alternative quantization with the left-right asymmetric vacuum used here, the level-matching condition gives a bound on the allowed levels. More explicitly, the only allowed levels are $(N,\bar{N}) = (1,1), (2,0), (0,2)$. As in the conventional quantization, we have to further impose the level one Virasoro constraint and the gauge equivalence due to spurious states. Then the physical spectrum is summarized in the following table.

\begin{table}[htb]
\begin{tabular}[t]{lccccc}
\hline
$N$ & $\bar{N}$ & $M^2$ & states & gauge condition & norm \\ \hline
$1$ & $1$ & $0$ & $e_{\mu\nu} \alpha^\mu_{-1}\bar{\alpha}^\nu_{+1}|0,k\rangle $ & $k^\mu e_{\mu\nu} = k^\nu e_{\mu\nu}=0$ & $+1$ \\
$2$ & $0$ & $+\frac{4}{\alpha'}$ & $A_{\mu\nu}\alpha^{\mu}_{-1}\alpha^\nu_{-1}|0,k\rangle$  & $k^{\mu}A_{\mu\nu} = A^\mu_{\ \mu}=0$  & $-1$ \\
$0$ & $2$ & $-\frac{4}{\alpha'}$ & $\bar{A}_{\mu\nu} \bar{\alpha}^\mu_{+1}\bar{\alpha}^\nu_{+1} |0,k\rangle $ &  $k^\mu\bar{A}_{\mu\nu} = \bar{A}^\mu_{\ \mu} =0$ & $-1$ \\ \hline
\end{tabular}
 \caption{Spectrum of bosonic strings with alternative vacuum choice.}
  \label{tb1}

\end{table}

The physical states with $(N,\bar{N}) = (1,1) $ contain graviton, Kalb-Ramond antisymmetric tensor and dilaton as in the conventional closed string theory. The physical states with $(N,\bar{N}) = (2,0) , (0,2)$ are interpreted as Fierz-Pauli symmetric tensors. The Fierz-Pauli symmetric tensor constructed out of the left-moving mode has negative mass squared while the one from the right-moving mode has positive mass squared. We also note that we do not have scalar tachyon in the physical state unlike in the conventional quantization of the closed string theory.

We have a remark on the norm of physical states. 
Here, following \cite{Lee:2017utr}, we have chosen 
\begin{align}
\langle  0 | 0 \rangle  < 0  
\end{align}
so that the norm of the physical gravity states is positive. We, however, note that this choice demands that the norm of physical Fierz-Pauli symmetric tensor fields is negative.

\section{Alternative quantization of R-NS string theory}
R-NS string theory is formulated by introducing world-sheet supersymmetry. For each bosonic field $X^\mu(\tau,\sigma)$, we introduce a fermionic field $\psi^\mu(\tau,\sigma)$ as a superpartner. Here $\psi^\mu(\tau, \sigma)$ is a two-component Majorana spinor on the world-sheet, which transforms as a Lorentz vector in the target space-time. Throughout the paper, we assume that the target space is $D$ dimensional flat Minkowski space-time. In the superconformal gauge, the supersymmetrized Polyakov action is given by\footnote{In this section except for the final mass formulae, we use $2\alpha' = 1$ \cite{Becker:2007zj}.}
\begin{align}
S = -\frac{1}{2\pi} \int d^2\sigma (\partial_\alpha X_\mu \partial^\alpha X^\mu + \bar{\psi}^\mu \rho^\alpha \partial_{\alpha} \psi_\mu ) \ ,  \label{RNS}
\end{align}
where $\rho^\alpha$ is a two-dimensional Dirac matrix satisfying the two dimensional Clifford algebra
\begin{align}
\{ \rho^\alpha, \rho^\beta \} = 2 \eta^{\alpha \beta} \ . 
\end{align}
One concrete realization of the Dirac matrix that we use is
\begin{align}
\rho^0 = \begin{pmatrix}
0 & -1 \\
1 & 0 \\
\end{pmatrix}  \ , \ \ \rho^1 = 
    \begin{pmatrix}
      0 & 1  \\
      1 & 0  \\
    \end{pmatrix}  \ . 
\end{align}
The Dirac conjugation is defined by $\bar{\psi} = \psi^\dagger i\rho^0$. 

With the explicit use of the component spinor $(A=\pm)$
\begin{align}
\psi^\mu_A = \left(
    \begin{array}{cc}
      \psi_-^\mu  \cr
      \psi_+^\mu 
    \end{array}
\right) \ , 
\end{align}
the fermionic part of the action becomes
\begin{align}
S_f = \frac{i}{\pi} \int d^2\sigma (\psi_- \cdot \partial_+ \psi_- + \psi_+ \cdot \partial_- \psi_+) \ ,  
\end{align}
from which we can deduce the equations of motion 
\begin{align}
\partial_+ \psi_-^\mu  = 0 \ , \ \ \partial_- \psi_+^\mu = 0 \ 
\end{align}
so that $\psi_-^\mu$ is a right-moving Majorana-Weyl spinor and $\psi_+^\mu$ is a left-moving Majorana-Weyl spinor.

\subsection{Energy-momentum tensor and supercurrent}
The action \eqref{RNS} is invariant under the supersymmetry transformation
\begin{align}
\delta X^\mu = \bar{\epsilon} \psi^\mu \ , \ \ \delta \psi^\mu =  \rho^\alpha \partial_\alpha X^\mu \epsilon \ , 
\end{align}
where $\epsilon$ is a Majorana spinor whose components are infinitesimal Grassmann numbers. Noether's theorem implies the existence of conserved current for the worldsheet supersymmetry. They are the energy-momentum tensor and the supercurrent. The energy-momentum tensor is given by
\begin{align}
T_{\alpha \beta} = \partial_\alpha X^\mu \partial_\beta X_\mu - \frac{1}{{2}}\eta_{\alpha\beta} \partial_\gamma X^\mu \partial^\gamma X_\mu + \frac{1}{4} \bar{\psi}^\mu \rho_\alpha \partial_\beta \psi_\mu + \frac{1}{4} \bar{\psi}^\mu \rho_\beta \partial_\alpha \psi_\mu -\frac{1}{4}\eta_{\alpha\beta}  \bar{\psi}^\mu \rho_\gamma \partial^\gamma \psi_\mu \ . 
\end{align}
Similarly, the supercurrent is given by
\begin{align}
J^\alpha_A = -\frac{1}{2} (\rho^\beta \rho^\alpha \psi_\mu)_A \partial_\beta X^\mu \ .
\end{align}
The existence of superconformal symmetry implies that the both are traceless: $T^{\alpha}_{\ \alpha} = \rho_\alpha J^\alpha = 0$.

In the light-cone coordinate, non-zero components of the energy-momentum tensor and the supercurrent are 
\begin{align}
T_{++} &= \partial_+ X_\mu \partial_+ X^\mu + \frac{i}{2} \psi^\mu_+ \partial_+ \psi_{+ \mu} \ , \ \ T_{--} = \partial_- X_\mu \partial_- X^\mu + \frac{i}{2} \psi_-^\mu \partial_- \psi_{-\mu} \ ,  \cr
J_+ &= \psi_+^\mu \partial_+ X_\mu \ , \ \ J_- = \psi_-^\mu \partial_- X_\mu \ . 
\end{align}
From the equations of motion of $X^\mu$ and $\psi^\mu$, we can show that they are conserved
\begin{align}
\partial_- T_{++} &= \partial_+ T_{--} = 0 \cr
\partial_- J_+ & = \partial_+ J_- = 0 \ . 
\end{align}

In the Polyakov formulation of the R-NS string theory, we have to impose the super Virasoro constraint from the variation of worldsheet metric and gravitino:
\begin{align}
J_+ = J_- = T_{++} = T_{--} = 0 \ . 
\end{align}
We will study the interpretation of these constraints in the alternative quantization of the R-NS string theory later.

\subsection{Mode expansion} 
The fermions in R-NS string theory admit two boundary conditions
\begin{align}
\psi_+^\mu (\sigma) = \pm \psi^\mu_+(\sigma + 2\pi) \cr
\psi_-^\mu (\sigma) = \pm \psi_-^\mu (\sigma + 2\pi) \ ,
\end{align}
which are compatible with the action  \eqref{RNS} and the worldsheet supersymmetry. For each mode, we can choose the sign independently and the case with $+$ is called Ramond (R) boundary condition and the case with $-$ is called Neveu-Schwarz (NS) boundary condition. In the R-sector, we have the mode expansion
\begin{align}
\psi_-^\mu (\tau,\sigma) = \sum_{n \in \mathbb{Z}} d_n^\mu e^{-2i n (\tau-\sigma)} \ ,  \ \
\psi_+^\mu (\tau,\sigma) = \sum_{n \in \mathbb{Z}} \bar{d}_n^\mu e^{-2i n (\tau+\sigma)}
\end{align}
 while in the NS-sector, we have the mode expansion
\begin{align}
\psi_-^\mu (\tau,\sigma) = \sum_{r \in \mathbb{Z} + \frac{1}{2} } b_r^\mu e^{-2i r (\tau-\sigma)} \ ,  \ \
\psi_+^\mu (\tau,\sigma) = \sum_{r \in \mathbb{Z}+ \frac{1}{2}} \bar{b}_r^\mu e^{-2i r (\tau+\sigma)} \ .
\end{align}

The mode expansion of the right-moving energy-momentum tensor and the supercurrent is given by
\begin{align}
T_{++} &= \sum_{n \in \mathbb{Z}} (L_n^b + L_n^{f_{\text{NS}}}) e^{-in(\tau -\sigma)} \ ,  \ \  J_{+} = \sum_{r \in \mathbb{Z}+\frac{1}{2}} G_r^{-ir(\tau-\sigma)} 
\end{align}
for the NS-sector, and 
\begin{align}
T_{++} &= \sum_{n \in \mathbb{Z}} (L_n^b + L_n^{f_{\text{R}}}) e^{-in(\tau -\sigma)} \ , \ \  J_{+} = \sum_{n \in \mathbb{Z}} F_n^{-in(\tau-\sigma)} 
\end{align}
for the R-sector. Here the super Virasoro generators are computed as
\begin{align}
L_m^b = \frac{1}{2} \sum_{n \in \mathbb{Z}} \alpha_{-n} \cdot \alpha_{m+n} \cr
L_m^{f_{\text{NS}}} = \frac{1}{2} \sum_{r \in \mathbb{Z}+\frac{1}{2}} (r+\frac{m}{2}) b_{-r} \cdot b_{m+r} \ , \ \ L_m^{f_{\text{R}}} = \frac{1}{2} \sum_{n \in \mathbb{Z}} (n+\frac{m}{2}) d_{-n} \cdot d_{m+n} \cr
F_m = \sum_{n \in \mathbb{Z}} \alpha_{-n} \cdot d_{m+n} \ , \ \ G_r = \sum_{n\in \mathbb{Z}} \alpha_{-n}\cdot b_{r+n}  
\end{align}
for $m \in \mathbb{Z} (\neq 0)$, $r\in \mathbb{Z} +\frac{1}{2}$. For the zero mode, we define them with the explicit normal ordering as
\begin{align}
L_0^b + L_0^{f_{\text{NS}}} &= \frac{1}{2}\alpha_0 \cdot \alpha_0 + \sum_{n=1}^\infty \alpha_{-n} \cdot \alpha_n + \sum_{r= \frac{1}{2} }^\infty r b_{-r} \cdot b_r  \cr
L_0^b + L_0^{f_{\text{R}}} &= \frac{1}{2}\alpha_0 \cdot \alpha_0 + \sum_{n=1}^\infty \alpha_{-n} \cdot \alpha_n + \sum_{n=1}^\infty n d_{-n} \cdot d_n \ . 
\end{align}

They satisfy the super Virasoro algebra
\begin{align}
[L_m, L_n] =  (m-n) L_{m+n} + \frac{D}{8}m (m^2-1) \delta_{m+n,0} \cr
[L_m, G_r] = (\frac{m}{2} - r) G_{m+r} \cr
\{ G_r, G_s \} = 2 L_{r+s} + \frac{D}{2} (r^2 -\frac{1}{4}) \delta_{r+s,0} 
\end{align}
for the NS-sector and
\begin{align}
[L_m, L_n] =  (m-n) L_{m+n} + \frac{D}{8}m^3 \delta_{m+n,0} \cr
[L_m, F_n] = (\frac{m}{2} - n) F_{m+n} \cr
\{ F_m, F_n \} = 2 L_{m+n} + \frac{D}{2}m^2 \delta_{m+n,0} 
\end{align}
for the R-sector.

We will come back to the left-moving mode in the following subsection.

\subsection{Alternative vacuum choice}

We have already discussed the alternative quantization of $X^\mu$ in section 2, so we will mostly focus on fermions.\footnote{The alternative vacuum choice for the R-NS string theory was briefly discussed in  \cite{Casali:2016atr}\cite{Casali:2017mss} in the context of the ambitwister string theory. Our results will agree with theirs.}
 The canonical commutation relation for the fermions is defined as 
\begin{align}
\{d_m^\mu, d_n^\nu\} &= \{\bar{d}_m^\mu ,\bar{d}_n^\nu \} = \delta_{m+n} \eta^{\mu\nu} \cr
\{b_r^\mu, b_s^\nu\} &= \{\bar{b}_r^\mu , \bar{b}_s^\nu \} = \delta_{r+s} \eta^{\mu\nu} \ . 
\end{align}

For the right-moving mode, we use the same vacuum as in the conventional R-NS string theory
\begin{align}
\alpha_m^\mu |0,k\rangle_{\text{R}} &= d_m^\mu |0, k \rangle_{\text{R}} = 0  \ (m > 0) \cr
\alpha_m^\mu |0,k\rangle_{\text{NS}} &= b_r^\mu |0, k \rangle_{\text{NS}} = 0  \ (m, r> 0) \ , \end{align}
while for the left-moving mode, we choose the asymmetric one \cite{Lee:2017utr}\cite{Casali:2017mss}:
\begin{align}
{}_{\text{R}} \overline{\langle 0,k |} \bar{\alpha}^\mu_m & = {}_{\text{R}}\overline{\langle 0, k|} \bar{d}_m^\mu  = 0  \ (m > 0) \cr
{}_{\text{NS}}\overline{\langle  0,k  |}\bar{\alpha}_m^\mu &= {}_{\text{NS}} \overline{\langle 0, k|} \bar{b}_r^\mu = 0  \ (m, r> 0) \ .
\end{align}
Note that the vacua in the R-sector in the right-moving sector as well as in the left-moving sector are degenerate due to the zero modes $d_0^\mu$ and $\bar{d}_0^\mu$. By identifying the anti-commutation relation with the ten-dimensional Clifford algebra, degenerate ground states form a spinor representation of the ten-dimensional Lorentz group as in the conventional R-NS string theory. 

In the left-moving sector, the super Virasoro generators are given by
\begin{align}
\bar{L}_m^b = \frac{1}{2} \sum_{n \in \mathbb{Z}}  \bar{\alpha}_{-n} \cdot \bar{\alpha}_{m+n} \cr
\bar{L}_m^{f_{\text{NS}}} = \frac{1}{2} \sum_{r \in \mathbb{Z}+\frac{1}{2}}^\infty (r+\frac{m}{2}) \bar{b}_{-r} \cdot \bar{b}_{m+r} \ , \ \ \bar{L}_m^{f_{\text{R}}} = \frac{1}{2} \sum_{n \in \mathbb{Z}} (n+\frac{m}{2}) \bar{d}_{-n} \cdot \bar{d}_{m+n} \cr
\bar{F}_m = \sum_{n \in \mathbb{Z}} \bar{\alpha}_{-n} \cdot \bar{d}_{m+n} \ , \ \ \bar{G}_r = \sum_{n\in \mathbb{Z}} \bar{\alpha}_{-n}\cdot \bar{b}_{r+n}  
\end{align}
for $m \in \mathbb{Z} (\neq 0)$, $r\in \mathbb{Z} +\frac{1}{2}$. For the zero mode $\bar{L}_0^{\text{R}} = \bar{L}_0^b + \bar{L}_0^{f_{\text{R}}}$,  $\bar{L}_0^{\text{NS}} = \bar{L}_0^b + \bar{L}_0^{f_{\text{NS}}}$, we define them with the explicit normal ordering as
\begin{align}
\bar{L}_0^{\text{R}} &= \frac{1}{2} \alpha_0 \cdot \alpha_0 - \bar{N}_{\text{R}} \ , \ \ \bar{N}_{\text{R}} = -\left(\sum_{n=1}^\infty \bar{\alpha}_{n} \cdot \bar{\alpha}_{-n} - \sum_{n=1}^\infty n \bar{d}_{n} \cdot \bar{d}_{-n} \right) \cr
\bar{L}_0^{\text{NS}} &=  \frac{1}{2} \alpha_0 \cdot \alpha_0 - \bar{N}_{\text{NS}} \ , \ \ \bar{N}_{\text{NS}} = -\left(\sum_{n=1}^\infty \bar{\alpha}_{n} \cdot \bar{\alpha}_{-n} - \sum_{r=\frac{1}{2}}^\infty r \bar{b}_{r} \cdot \bar{b}_{-r} \right)  \ . 
\end{align}
They satisfy the super Virasoro algebra
\begin{align}
[\bar{L}_m, \bar{L}_n] =  (m-n) \bar{L}_{m+n} - \frac{D}{8}m (m^2-1) \delta_{m+n,0} \cr
[\bar{L}_m, \bar{G}_r] = (\frac{m}{2} - r) \bar{G}_{m+r} \cr
\{ \bar{G}_r, \bar{G}_s \} = 2 \bar{L}_{r+s} - \frac{D}{2} (r^2 -\frac{1}{4}) \delta_{r+s,0} 
\end{align}
for the NS-sector and
\begin{align}
[\bar{L}_m, \bar{L}_n] =  (m-n) \bar{L}_{m+n} - \frac{D}{8}m^3 \delta_{m+n,0} \cr
[\bar{L}_m, \bar{F}_n] = (\frac{m}{2} - n) \bar{F}_{m+n} \cr
\{ \bar{F}_m, \bar{F}_n \} = 2 \bar{L}_{m+n} - \frac{D}{2}m^2 \delta_{m+n,0} 
\end{align}
for the R-sector. {Note the sign difference in the central terms.}

We now postulate the super Virasoro constraint on the physical states in the alternative quantization of the R-NS string theory. On the right-moving sector, we demand
\begin{align}
G_r |\mathrm{phys} \rangle = 0  \ \  (r>0) \cr
L_m |\mathrm{phys} \rangle = 0 \ \ (m>0)  \cr
(L_0 - a_{\text{NS}}) |\mathrm{phys} \rangle = 0 
\end{align}
in the NS-sector, and 
\begin{align}
F_n |\mathrm{phys} \rangle = 0  \ \ (n \ge 0)  \cr
L_m |\mathrm{phys} \rangle = 0 \ \ (m>0)  \cr
(L_0 - a_{\text{R}}) |\mathrm{phys} \rangle = 0 
\end{align}
in the R-sector.
They are same as in the conventional R-NS string theory. As for the left-moving sector, we demand 
\begin{align}
\langle \mathrm{phys}| \bar{G}_r = 0 \ \ (r>0) \cr
\langle \mathrm{phys}| \bar{L}_m = 0 \ \ (m>0) \cr
\langle \mathrm{phys}| (\bar{L}_0 - \bar{a}_{\text{NS}}) = 0 
\end{align} 
in the NS-sector, and 
\begin{align}
\langle \mathrm{phys}| \bar{F}_n = 0  \ \ (n \ge 0) \cr
\langle \mathrm{phys}| \bar{L}_m = 0 \ \ (m>0) \cr
\langle \mathrm{phys}| (\bar{L}_0 - \bar{a}_{\text{R}}) = 0  
\end{align}
in the R-sector. Here we have introduced the normal ordering constant $a_{\text{NS}}, a_{\text{R}}, \bar{a}_{\text{NS}}, \bar{a}_{\text{R}}$. 

We can determine the normal ordering constant as $a_{\text{R}} = 0 $, $a_{\text{NS}}=\frac{1}{2}$, $\bar{a}_{\text{R}} = 0$, $\bar{a}_{\text{NS}} = -\frac{1}{2}$ by requiring that the physical states include massless gravity states. This also fixes the space-time dimension $D=10$. 
In summary, we have the following list of the mass-shell condition and the level-matching condition for each sector.

In the  $(\text{NS},\text{NS})$ sector, we have
\begin{align}
M^2 &= \frac{4}{\alpha'}(N_{\text{NS}} - \frac{1}{2}) = \frac{4}{\alpha'}(-\bar{N}_{\text{NS}} + \frac{1}{2}) \cr
N_{\text{NS}} + \bar{N}_{\text{NS}} &= \frac{1}{2}+\frac{1}{2} = 1 \ . 
\end{align}
In the $(\text{NS},\text{R})$ sector, we have
\begin{align}
M^2 &= \frac{4}{\alpha'}(N_{\text{NS}} - \frac{1}{2}) = \frac{4}{\alpha'}(-\bar{N}_{\text{R}} + 0 ) \cr
N_{\text{NS}} + \bar{N}_{\text{R}} &= \frac{1}{2}+ 0 = \frac{1}{2} \ . 
\end{align}
In the $(\text{R},\text{NS})$ sector, we have
\begin{align}
M^2 &= \frac{4}{\alpha'}(N_{\text{R}} - 0) = \frac{4}{\alpha'}(-\bar{N}_{\text{NS}} + \frac{1}{2}) \cr
N_{\text{R}} + \bar{N}_{\text{NS}} &= 0 +\frac{1}{2} = \frac{1}{2} \ . 
\end{align}
In the  $(\text{R},\text{R})$ sector, we have
\begin{align}
M^2 &= \frac{4}{\alpha'}(N_{\text{R}} - 0) = \frac{4}{\alpha'}(-\bar{N}_{\text{R}} + 0) \cr
N_{\text{R}} + \bar{N}_{\text{R}} &= 0 + 0  = 0 \ . 
\end{align}

\subsection{Physical spectrum and GSO projection}
We now analyze the physical states in the alternative quantization of the R-NS string theory after imposing the level-matching condition. As in the alternative quantization of the bosonic string theory reviewed in section 2, it is crucial to observe that we have only a finite number of propagating degrees of freedom unlike in the conventional quantization of the R-NS string theory.

In the $(\text{NS},\text{NS})$ sector, the level matching condition is 
\begin{align}
N_{\text{NS}} + \bar{N}_{\text{NS}} = \frac{1}{2} + \frac{1}{2} = 1 \ .
\end{align}
Consequently, the allowed spectrum are $(N_{\text{NS}},\bar{N}_{\text{NS}}) = (1,0),(0,1), (\frac{1}{2},\frac{1}{2})$. The explicit states are
\begin{align}
(1,0): \ \  \alpha^\mu_{-1}|0,k\rangle_{\text{NS}} \overline{|0,k\rangle}_{\text{NS}} \  ,  \ \ M^2 = \frac{2}{\alpha'} \cr
(0,1): \ \ \bar{\alpha}^\mu_{+1}|0,k\rangle_{\text{NS}} \overline{|0,k\rangle}_{\text{NS}} \ , \ \ M^2 = -\frac{2}{\alpha'} \cr
(\frac{1}{2},\frac{1}{2}): \ \  b^\mu_{-\frac{1}{2}}\bar{b}^\nu_{+\frac{1}{2}} |0,k\rangle_{\text{NS}} \overline{|0,k\rangle}_{\text{NS}} \  , \ \ M^2 = 0 \ .
\end{align}
Here the ground states with bar are for the left-moving mode. After imposing the super Virasoro constraint, the physical states describe two massive vectors with positive and negative mass squared,  massless graviton, massless Kalb-Ramond antisymmetric tensor and massless dilaton. Note that it does not include a scalar tachyon unlike in the conventional quantization.

The norm of each states (with transverse polarization) are
\begin{align}
(1,0): {}_{\text{NS}}{ \langle 0,k|} {}_{\text{NS}}\overline{ \langle 0,k|} \alpha^I_{+1} \alpha^J_{-1} \overline{|0,k\rangle}_{\text{NS}}{|0,k\rangle}_{\text{NS}}  \sim + \delta^{IJ} {}_{\text{NS}}{ \langle 0,k} {|0,k\rangle}_{\text{NS}} {}_{\text{NS}}\overline{ \langle 0,k|}  \overline{0,k\rangle}_{\text{NS}} 
\cr
(0,1):   {}_{\text{NS}}{ \langle 0,k|}{}_{\text{NS}}\overline{ \langle 0,k|}  \bar{\alpha}^I_{-1} \bar{\alpha}_{+1}^J \overline{|0,k\rangle}_{\text{NS}} {|0,k\rangle}_{\text{NS}} \sim - \delta^{IJ} {}_{\text{NS}}{ \langle 0,k} {|0,k\rangle}_{\text{NS}} {}_{\text{NS}}\overline{ \langle 0,k|} \overline{0,k\rangle}_{\text{NS}} \cr
(\frac{1}{2},\frac{1}{2}):  {}_{\text{NS}}{ \langle 0,k|} {}_{\text{NS}}\overline{ \langle 0,k|} b^I_{+\frac{1}{2}} \bar{b}^{J}_{-\frac{1}{2}} b^{K}_{-\frac{1}{2}} \bar{b}^L_{+\frac{1}{2}} \overline{|0,k\rangle}_{\text{NS}} {|0,k\rangle}_{\text{NS}}  \sim + \delta^{IK}\delta^{JL}  {}_{\text{NS}}{ \langle 0,k} {|0,k\rangle}_{\text{NS}} {}_{\text{NS}}\overline{ \langle 0,k|} \overline{0,k\rangle}_{\text{NS}}
\end{align}

In the $(\text{NS},\text{R})$ sector, the level matching condition is 
\begin{align}
N_{\text{NS}} + \bar{N}_{\text{R}} = \frac{1}{2} + 0 = \frac{1}{2} \ .
\end{align}
Consequently, the allowed spectrum is $(N_{\text{NS}},\bar{N}_{\text{R}}) = (\frac{1}{2},0)$.  The explicit state is 
\begin{align}
(\frac{1}{2},0): \ \ b^\mu_{-\frac{1}{2}}|0,k\rangle_{\text{NS}} \overline{|0,k\rangle}_{\text{R}} \ ,  \ \ M^2 = 0 \ .
\end{align}
The degeneracy of the R sector vacuum, which is implicit, means that the left-moving sector transforms as (two) ten dimensional spinors (with both chiralities).
It is same as the massless spectrum of conventional quantization of the R-NS string theory. The norm is
\begin{align}
{}_{\text{NS}} \overline{\langle 0,k|} {}_{\text{R}} \langle 0,k| \bar{b}_{-\frac{1}{2}}^I \bar{b}^{J}_{+\frac{1}{2}}|0,k\rangle_{\text{R}} \overline{ |0,k\rangle}_{\text{NS}} \sim + \delta^{IJ} {}_{\text{NS}} \overline{\langle 0,k}| \overline{0,k \rangle}_{\text{NS}} {}_{\text{R}} \langle 0, k|0,k \rangle_{\text{R}} \ . 
\end{align}

In the $(\text{R},\text{NS})$ sector, the level matching condition is 
\begin{align}
N_{\text{R}} + \bar{N}_{\text{NS}} = 0 + \frac{1}{2}  = \frac{1}{2} \ .
\end{align}
Consequently, the allowed spectrum is $(N_{\text{R}},\bar{N}_{\text{NS}}) = (0, \frac{1}{2})$. The explicit state is 
\begin{align}
(0,\frac{1}{2}): \ \ \bar{b}^\mu_{+\frac{1}{2}}|0,k\rangle_{\text{R}} \overline{|0,k\rangle}_{\text{NS}} \  ,  \ \ M^2 = 0 \ .
\end{align}
It is same as the massless spectrum of conventional quantization of the R-NS string theory.
The norm is
\begin{align}
{}_{\text{R}} \overline{\langle 0,k|} {}_{\text{NS}} \langle 0,k| {b}_{+\frac{1}{2}}^I {b}^{J}_{-\frac{1}{2}}|0,k\rangle_{\text{NS}} \overline{ |0,k\rangle}_{\text{R}} \sim + \delta^{IJ} {}_{\text{R}} \overline{\langle 0,k}| \overline{0,k \rangle}_{\text{R}} {}_{\text{NS}}\langle 0, k |0,k \rangle_{\text{NS}} \ . 
\end{align}

In the $(\text{R},\text{R})$ sector, the level matching condition is 
\begin{align}
N_{\text{R}} + \bar{N}_{\text{R}} = 0 + 0  = 0 \ .
\end{align}
Consequently, the allowed spectrum is $(N_{\text{R}},\bar{N}_{\text{R}}) = (0, 0)$. The explicit state is 
\begin{align}
(0, 0 ): \ \ |0,k\rangle_{\text{R}} \overline{|0,k\rangle}_{\text{R}}  \ ,  \ \ M^2 = 0 \ .
\end{align}
Is is same  as the massless spectrum of conventional quantization of the R-NS string theory. The norm is 
\begin{align}
({}_{\text{R}} \overline{\langle 0,k|} {}_{\text{R}} \langle 0,k|)(|0,k\rangle_{\text{R}} \overline{ |0,k\rangle}_{\text{R}}) \sim + {}_{\text{R}} \overline{\langle 0,k}| \overline{0,k \rangle}_{\text{R}} {}_{\text{R}}\langle 0, k |0,k \rangle_{\text{R}} . 
\end{align}

The results are summarized in table 2. Here the norm of the ground state is chosen so that the gravity sector becomes massless, but this choice makes one massive vector be equipped with a negative norm.

\begin{table}[htb]
\begin{tabular}[t]{lccccc}
\hline
 & $N$ & $\bar{N}$ & $M^2$ & representation & norm \\ \hline
(NS,NS) & $\frac{1}{2}$ & $\frac{1}{2}$ & 0 & vector $\otimes$ vector &$+$ \\
(NS,NS) & $1$ & $0$ & $\frac{2}{\alpha'}$ & vector &$+$   \\
(NS,NS) & $0$ & $1$ & $-\frac{2}{\alpha'}$ & vector&$-$   \\
(NS,R) & $\frac{1}{2}$ & $0$ & $0$ & vector $\otimes$ spinor &$+$  \\
(R,NS) & $0$ & $\frac{1}{2}$ & $0$ & spinor $\otimes$ vector &$+$  \\
(R,R) & $0$ & $0$  & $0$  & spinor $\otimes$ spinor &$+$  \\ \hline
\end{tabular}
 \caption{Spectrum of R-NS strings with alternative vacuum choice.}
  \label{tb2}
\end{table}

In the conventional quantization of the R-NS string theory, we can impose the GSO projection to realize the target space supersymmetry. It seems reasonable to pursue the same idea in the alternative quantization of the R-NS string theory.

On the NS sector, we introduce the G-parity operator in the right-moving sector and the left-moving sector as
\begin{align}
G_{\text{NS}} &= (-1)^{\sum_{r=\frac{1}{2}}^\infty b_{-r} \cdot b_r + 1} \cr
\bar{G}_{\text{NS}} & =(-1)^{\sum_{r=\frac{1}{2}}^\infty \bar{b}_r \cdot \bar{b}_{-r} + 1}
\end{align}
and we declare that physical states satisfy the condition
\begin{align}
G_{\text{NS}} | \text{phys} \rangle &= |\text{phys} \rangle \cr
\langle \text{phys}| \bar{G}_{\text{NS}} &= \langle \text{phys}| \ . 
\end{align}

On the R sector, we need to prescribe the G-parity of the zero modes. We define the G-parity operator as
\begin{align}
G_{\text{R}} &= \Gamma_{11} (-1)^{\sum_{n=1}^\infty d_{-n} \cdot d_{n}} \cr
\bar{G}_{\text{R}} &= (-1)^{\sum_{n=1}^\infty \bar{d}_{n} \cdot \bar{d}_{-n}} \Gamma_{11} \ ,
\end{align}
where $\Gamma_{11} = \Gamma_0\Gamma_1 \cdots \Gamma_9$ is the chirality operator in ten dimensions by identifying $d_{0}^\mu$ with the Gamma matrix. We thus identify the zero modes of the R sector as a space-time spinor and we decompose the right-moving as well as left-moving ground states into the chiraliy eigenstates as
\begin{align}
|0,k\rangle_{\text{R}} = \left(
    \begin{array}{cc}
      | + \rangle  \cr
      |- \rangle
    \end{array}
\right) \ ,   \ \ \ 
{}_{\text{R}} \overline{ \langle 0,k |} = \left(
    \begin{array}{cc}
      \overline {\langle + |}  \cr
     \overline{\langle - |}
    \end{array}
\right) \ , 
\end{align}
with
\begin{align}
G_{\text{R}}|+ \rangle = |+ \rangle \ , \ \ G_{\text{R}}|-\rangle = -|- \rangle \ , \ \  \overline{ \langle +|} \bar{G}_{\text{R}} = \overline{\langle +|} \ , \ \ \ \overline{\langle - |} \bar{G}_{\text{R}}  = - \overline{\langle -|} \ . 
\end{align}

As in the conventional quantization of the R-NS string theory, in the alternative quantization, we can choose the same or opposite sign for the states that we would like to keep in the left-moving mode of the R-sector (as compared with the right-moving mode). This results in theories similar to type IIA and typie IIB superstring theory (in the conventional quantization), which we call alt IIA theory and alt IIB theory.\footnote{We could consider alt 0A and alt 0B theory in analogy to type 0A and type 0B string theory by keeping all the $(\text{NS},\text{NS})$ sector and half of the $(\text{R},\text{R})$ sector. In these cases, we obtain different low-lying spectrums from type 0A and type 0B string theory. In particular, we do not have scalar tachyon, but we have two massive vectors with positive and negative mass squared. A possible lack of modular invariance with an alternative quantization \cite{Lee:2017utr}, however, might make such choices less grounded.}

Alt IIA theory is obtained by choosing the  opposite sign for the G-parity in the R ground states of the left-moving and right-moving modes. After the projection, physical states are
\begin{align}
b^{\mu}_{-\frac{1}{2}} \bar{b}^\nu_{+\frac{1}{2}} |0 \rangle_{\text{NS}} \overline{|0 \rangle}_{\text{NS}} \cr 
b^\mu_{-\frac{1}{2}} |0\rangle_{\text{NS}} \overline{|-\rangle}_{\text{R}} \cr
\bar{b}^\mu_{+\frac{1}{2}} |+\rangle_{\text{R}} \overline{| 0 \rangle}_{\text{NS}} \cr
|+ \rangle_{\text{R}} \overline{|- \rangle}_{\text{R}}  \ ,
\end{align}
all of which are massless $k^2=0$.\footnote{We are implicit about the extra conditions coming from the level one super Virasoro constraint that gives the transverse condition on the polarization and the Dirac equation from $F_0$ and $\bar{F}_0$.}

Alt IIB theory is obtained by choosing the same sign for the G-parity in the R ground states of the left-moving and right-moving modes. After the projection, physical states are
\begin{align}
b^{\mu}_{-\frac{1}{2}} \bar{b}^\nu_{+\frac{1}{2}} |0 \rangle_{\text{NS}} \overline{|0 \rangle}_{\text{NS}} \cr
b^\mu_{-\frac{1}{2}} |0\rangle_{\text{NS}} \overline{|+\rangle}_{\text{R}} \cr
\bar{b}^\mu_{+\frac{1}{2}} |+\rangle_{\text{R}} \overline{| 0 \rangle}_{\text{NS}} \cr
|+ \rangle_{\text{R}} \overline{|+ \rangle}_{\text{R}}  \ .
\end{align}

In summary, we obtain the following physical states. Both alt IIA theory and alt IIB theory have the same spectrum in the $(\text{NS},\text{NS})$ sector. They are massless dilaton, Kalb-Ramond antisymmetric tensor and graviton. In the $(\text{R},\text{NS})$ and $(\text{NS},\text{R})$ sector, they have massless dilatino and gravitino. In alt IIA theory, two gravitino (dilatino) have the opposite chirality while in alt IIB theory, they have the same chirality. In the $(\text{R},\text{R})$ sector, in alt IIA theory, it has massless 1-form and 3-form gauge fields  while in alt IIB theory, it has massless 0-form, 2-form and self-dual 4-form gauge fields. We see that they actually coincide with the field contents of type IIA and type IIB supergravity in ten dimensions, or keeping only massless modes of the corresponding superstring theories. 
Thus the propagating degrees of freedom of the alternative quantization of the R-NS string theory after GSO projection have no ghost states and they are anomaly free. 

\section{Discussions}
In this paper, we investigated the alternative quantization of the R-NS string theory based on the left-right asymmetric choice of vacua. One of the most interesting properties of the resulting theory is that it has only a finite number of propagating degrees of freedom. After imposing the supersymmetric GSO projection, the spectrum coincides with type IIA or type IIB supergravity without any massive excitations.\footnote{Thus, we may conclude that the spectrum of the alternative quantization of the R-NS string theory with the supersymmetric GSO projection is the same as that of the ambitwister string theory \cite{Mason:2013sva}\cite{Adamo:2014wea} (with infinite string tension). A similar construction of the chiral superstring theory was studied in \cite{Leite:2016fno}. Some of the results of this paper should have appeared in \cite{Rey:2017}, but the authors are not aware of the contents.}

For a future direction to be pursued, it is important to study higher loop amplitudes to see if the theory is ultraviolet finite.\footnote{The alternative vacuum choice may not make sense in the open string sector due to its intrinsic asymmetry. Without D-branes, nonperturbative completion of the alternative quantization needs further investigations.} It is also important to understand the consistency of the quantization in non-trivial supergravity background. Even a circle compactification is non-trivial \cite{Casali:2017mss}\cite{Lee:2017crr} and may introduce negative norm states. In relation, quantization of heterotic string theory on the alternative vacuum may be of interest.

\end{document}